\documentclass[aps,prd,nofootinbib,showpacs,superscriptaddress,12pt]{revtex4}
\usepackage{txfonts}
\usepackage{graphicx}
\usepackage{dcolumn}
\usepackage{bm}
\usepackage{amssymb}
\usepackage{latexsym}

\begin{document}

\title{\bf Cosmological Constraint and Analysis on Holographic Dark Energy Model Characterized by the Conformal-age-like Length}
\author{Zhuo-Peng~Huang}
\email[Electronic address: ]{zphuang@nudt.edu.cn}
\affiliation{Department of Physics, National University of Defense Technology, Hunan 410073, China}
\affiliation{State Key Laboratory of Theoretical Physics (SKLTP)\\
Kavli Institute for Theoretical Physics China (KITPC) \\
Institute of Theoretical Physics, Chinese Academy of Sciences, Beijing 100190, China}
\author{Yue-Liang~Wu}
\email[Electronic address: ]{ylwu@itp.ac.cn}
\affiliation{State Key Laboratory of Theoretical Physics (SKLTP)\\
Kavli Institute for Theoretical Physics China (KITPC) \\
Institute of Theoretical Physics, Chinese Academy of Sciences, Beijing 100190, China}

\date{\today}

\begin{abstract}

We present a best-fit analysis on the single-parameter holographic dark energy model characterized by the conformal-age-like length, $L=
\frac{1}{a^4(t)}\int_0^tdt'~a^3(t') $. Based on the Union2 compilation of 557 supernova Ia data, the baryon acoustic oscillation results from the SDSS DR7 and the
cosmic microwave background radiation data from the WMAP7, we show that the model gives the minimal $\chi^2_{min}=546.273$, which is comparable to $\chi^2_{\Lambda
{\rm CDM}}=544.616$ for the $\Lambda$CDM model. The single parameter $d$ concerned in the model is found to be $d=0.232\pm 0.006\pm 0.009$.
Since the fractional density of dark energy $\Omega_{de}\sim d^2a^2$ at $a \ll 1$, the fraction of dark energy is naturally negligible in the early universe,
$\Omega_{de} \ll 1$ at $a \ll 1$. The resulting constraints on the present fractional energy density of matter and the equation of state are
$\Omega_{m0}=0.286^{+0.019}_{-0.018}~ ^{+0.032}_{-0.028}$ and $w_{de0}=-1.240^{+0.027}_{-0.027}~ ^{+0.045}_{-0.044}$ respectively. The model leads to a slightly
larger fraction of matter comparing to the $\Lambda$CDM model. We also provide a systematic analysis on the cosmic evolutions of the fractional energy density of
dark energy, the equation of state of dark energy, the deceleration parameter and the statefinder. It is noticed that the equation of state crosses from $w_{de}>-1$
to $w_{de}<-1$, the universe transits from decelerated expansion ($q>0$) to accelerated expansion ($q<0$) recently, and the statefinder may serve as a sensitive
diagnostic to distinguish the CHDE model with the $\Lambda$CDM model.

\end{abstract}

\maketitle


\section{Introduction}

Observations of Type Ia supernovae (SNIa) \cite{Riess:1998cb, Perlmutter:1998np}, cosmic microwave background (CMB) \cite{Spergel:2003cb} and large scale structure
(LSS) \cite{Tegmark:2003ud} have complementarily established the present acceleration of the universe expansion. Within the framework of the general relativity, a
consistent picture has indicated that nearly three quarters of our universe consists of a mysterious negative pressure component named by dark energy, which is
responsible for the accelerated expansion. However, the nature of such an exotic energy component is still rather uncertain. The simplest candidate is a positive
cosmological constant. Although fitting the observations well, a cosmological constant, however, is plagued with the fine-tuning problem and the coincidence problem
\cite{Weinberg:1988cp}.

The holographic principle \cite{'tHooft:1993gx, Susskind:1994vu} indicates that the vacuum energy calculated in quantum field theory might take too many degrees of
freedom into consideration which results in the fine-tuning problem of the cosmological constant. In \cite{Cohen:1998zx}, the author suggested that due to the limit
set by the formation of a black hole, the ultraviolet (UV) cutoff $\Lambda_{\rm uv}$ in the effective field theory should be related to the infrared (IR) cutoff
$L$, i.e., in terms of the natural units,
 \begin{equation}
    L^3\Lambda_{\rm uv}^4 \lesssim LM_p^2  ~,
 \end{equation}
where $M_P$ is the reduced Planck constant $M_p^2=1/(8\pi G)$ with $G$ the Newton's~constant. This means that the effective theory describes all states of the
system, except those that have already collapsed to a black hole.  Such a dramatic depletion of quantum states leads to much small vacuum energy density,
 \begin{equation}
    \rho_{\rm vac} \sim \Lambda_{\rm uv}^4 \sim  M_p^2L^{-2}  ~.
 \end{equation}
If the IR cutoff $L$ is comparable to the current Hubble radius, the resulting $\rho_{\rm vac}$ requires no magnificent cancelation to be consistent with
observational bounds \cite{Cohen:1998zx}. Many interesting studies on holography and cosmology are conducted
\cite{Horava:2000tb,Thomas:2002pq,Fischler:1998st,Bousso:1999xy}. In \cite{Huang:2012xm}, we show that if the IR cutoff $L$ is characterized with the total comoving
horizon of the universe, the very large primordial part of the comoving horizon generated by the inflation of early universe might give some insights into the
cosmological constant and the coincidence problem. Plenty of alternative models ( for some reviews see
\cite{Peebles:2002gy,Padmanabhan:2002ji,Copeland:2006wr,Tsujikawa:2010sc, Li:2011sd}) have been proposed to provide the possible explanations for the recent cosmic
acceleration. Based on the holographic principle \cite{'tHooft:1993gx, Susskind:1994vu, Cohen:1998zx}, some interesting holographic dark energy models in which the
dark energy density is assumed to scale as $\rho_{de}\propto M_P^2L^{-2}$, were proposed and studied
\cite{Hsu:2004ri,Li:2004rb,Huang:2004ai,Gong:2004fq,Myung:2004ch,Pavon:2005yx,Wang:2005jx,Hu:2006ar,
Chen:2006qy,Li:2006ci,Setare:2006wh,Cai:2007us,Wei:2007ty,Gao:2007ep,Granda:2008dk,Granda:2008tm,Gong:2009dc,
Jamil:2009sq,Feng:2009hr,delCampo:2011jp,Gao:2011,Li:2009bn,Huang:2012nz}  by taking different choices of the characteristic length scale of the universe, $L$.
Especially, the age of the universe was chosen as the IR cutoff to build agegraphic dark energy model(ADE) \cite{Cai:2007us} . In order to avoid some internal
inconsistencies in the original model, a new version of this model (NADE) replacing the age of the universe by the conformal age of the universe \cite{Wei:2007ty}.
An interesting comparison of some holographic dark energy models in \cite{Li:2009bn} shows that the NADE seems not to be consistent with the cosmological
observations very well. Writing the four dimensional spacetime volume at the cosmic time $t$ as follows
 \begin{equation}
    \int d^3x \int _0^t dt' \sqrt{-g}=\left( a^3(t)\int d^3x \right) \cdot a(t) \cdot \left(  \frac{1}{a^4(t)}\int_0^t dt'~a^3(t')\right) \equiv V~ a(t)~ L ~,
 \end{equation}
where $V= a^3(t)\int d^3x $ is the physical space volume and $a(t)$ is the scale factor of the universe, and $L$ is defined to be
 \begin{equation}
    L=\frac{1}{a^4(t)}\int_0^t dt'~a^3(t') = \frac{1}{a^4(t)}\int \frac{dt'}{a(t')} ~a^4 (t')  \equiv \frac{1}{a^4(t)}\int  d\eta' ~a^4 (t')
 \end{equation}
which may be regarded as a conformal-age-like length scale of the universe. In \cite{Huang:2012nz}, such conformal-age-like length is proposed to be the
characteristic length scale of the universe to establish a holographic dark energy model (CHDE) which is similar to the new agegraphic dark energy model
\cite{Wei:2007ty}. The conformal-age-like length $\frac{1}{a^4(t)}\int_0^t dt'~a^3(t')$ is adopted rather than the age-like length $\frac{1}{a^3(t)}\int_0^t
dt'~a^3(t')$, this is because the model based on the age-like length seems to have the similar self-inconsistency to the agegraphic dark energy model(ADE)\cite{Cai:2007us}.

In this note, we are going to perform a best-fit analysis on the CHDE model by using the Union2 compilation of 557 supernova Ia (SNIa) data \cite{Amanullah:2010vv},
the baryon acoustic oscillation (BAO) results from the Sloan Digital Sky Survey data release 7 (SSDS DR7) \cite{Percival:2009xn} and the cosmic microwave background
radiation (CMB) data from the 7-yr Wilkinson Microwave Anisotropy Probe (WMAP7) \cite{Komatsu:2010fb}. We find that the best-fit results of CHDE model are
comparable to that of the $\Lambda$CDM model. Based on the observational constraints on the CHDE model, we also concentrate on the studies for the cosmic evolutions
of some interesting quantities within the CHDE model. The paper is organized as follows: in Sec.\,II, we briefly introduce the CHDE model; in Sec.\,III, we carry
out a best-fit analysis on the CHDE model; in Sec.\,IV, we study the cosmic evolutions of the fractional energy density of CHDE, the equation of state of CHDE, the
deceleration parameter and statefinder; our concluding remarks are given in Sec.\,V.

\section{Brief Outline on CHDE Model}

Rewriting the conformal-age-like parameter as
 \begin{equation}
    L=\frac{1}{a^4(t)}\int_0^t dt'~a^3(t')=\frac{1}{a^4(t)}\int_0^a a'^3\frac{da'}{H'a'}~,  \label{l}
 \end{equation}
with $H\equiv \dot{a}/a$ the Hubble parameter and $``\cdot"$ denoting the derivative respect to cosmic time $t$, we then define the holographic dark energy with the density parameterized by the characteristic length scale $L$ of the universe as follows
 \begin{equation}
    \rho_{de}=3d^2 M_p^2 L^{-2}~, \label{rho}
 \end{equation}
where $d$ is a positive constant parameter. Correspondingly, the fractional energy density is given by the characteristic length scale $L$ of the universe
 \begin{equation}
    \Omega_{de}=\frac{\rho_{de}}{3M_p^2H^2}=\frac{d^2}{H^2L^2} ~. \label{frho}
 \end{equation}

Considering a flat FRW universe containing matter, radiation and CHDE,  we have the Friedmann equation
 \begin{equation}
    3M_p^2H^2= \rho_{m} + \rho_{r} + \rho_{de} ~.                          \label{fri0}
 \end{equation}
When each component is conservative respectively, we get the equation for the density
 \begin{equation}
    \dot{\rho}_{i}+3H(1+w_{i})\rho_{i}=0  ~ \label{ceq}
 \end{equation}
with $i=m$, $r$ and $de$. Combining with Eqs.(\ref{l}-\ref{frho}), the EoS of dark energy is given by \cite{Huang:2012nz}
 \begin{equation}
    w_{de}=-1 - \frac83 + \frac2{3d}\frac{\sqrt{\Omega_{de}}}a~.\label{wde}
 \end{equation}
The conservations of matter and radiation result in $\rho_{m}=\rho_{m0}a^{-3}=\Omega_{m0} 3M_p^2H_0^2 a^{-3}$ and $\rho_{r}=\rho_{r0}a^{-4}=\Omega_{r0} 3M_p^2H_0^2
a^{-4}$ respectively, where the fractional energy densities are defined as $\Omega_i=\rho_i/\rho_c$ for $i=m$ and $r$, and $\rho_c=3M_p^2H^2$ is the critical energy
density. Note that we have set the present scale factor of the universe $a_0=1$ and the subscript ''0'' always indicates the present values of the corresponding
quantities. Thus the Friedmann equation Eq.(\ref{fri0}) can be rewritten as
 \begin{equation}
    H^2= \Omega_{m0} H_0^2 a^{-3} + \Omega_{r0} H_0^2a^{-4} + \Omega_{de}H^2 ~.                          \label{fri}
 \end{equation}
Defining $ r_0=\Omega_{r0} / \Omega_{m0} $, we have
 \begin{equation}
     \Omega_r(a)=\frac{\rho_r}{3M_p^2H^2}=\frac{r_0}{a+r_0}(1-\Omega_{de}(a))~.   \label{or2}
 \end{equation}
 and
 \begin{equation}
     \Omega_m(a)=\frac{\rho_m}{3M_p^2H^2}=\frac{a}{a+r_0}(1-\Omega_{de}(a) )~.    \label{or1}
 \end{equation}
From the Friedmann equation Eq.(\ref{fri}), we have
 \begin{equation}
    \frac1{Ha}=\frac1{H_0\sqrt{\Omega_{m0}}}\sqrt{a(1-\Omega_{de})}\sqrt{\frac 1{1+r_0/a}}~.\label{ha}
 \end{equation}
Referring to Eqs.(\ref{l}) and (\ref{frho}), one gets
 \begin{equation}
    \int_0^a a'^3\frac{da'}{H'a'}=\frac{a^5 d }{\sqrt{\Omega_{de}}Ha} ~.
 \end{equation}
Substituting Eq. (\ref{ha}) into above equation and taking derivative with respect to $a$ in both sides, we obtain the differential equation of motion for
$\Omega_{de}$ as \cite{Huang:2012nz}
 \begin{equation}
    \frac{d\Omega_{de}}{da}=\frac{\Omega_{de}}{a}(1-\Omega_{de})\left(11+\frac{r_0}{a+r_0}-\frac2d\frac{\sqrt{\Omega_{de}}}a\right) ~. \label{ode}
 \end{equation}
Under the limit $1-\Omega_{de} \simeq 1$ at $a \ll 1$, we arrive at the solution from the differential equation (\ref{ode})
 \begin{equation}
    \Omega_{de}\simeq\frac{d^2}{4}\left( 9+\frac{r_0}{a+r_0} \right)^2 a^2 ~ \label{frhom}
 \end{equation}
which is a good approximation and consistent with the conditions $a \ll 1$ and $1-\Omega_{de} \simeq 1$.

It is noticed that the above solution Eq.(\ref{frhom}) reduces to the corresponding self-consistent approximation in radiation-dominated epoch ($r_0 \to 1$) and in
matter-dominated epoch ($r_0 \to 0$) respectively \cite{Huang:2012nz}. Substituting Eq.(\ref{frhom}) into Eq.(\ref{wde}), we yield the EoS of the dark energy in
ambient-constituent-dominated epoch. Particularly, the EoS of dark energy is $w_{de}=-\frac13$ in the radiation-dominated epoch, and then transits to
$w_{de}=-\frac23$ in matter-dominated epoch. Referring to Eq.(\ref{wde}), the EoS of dark energy $w_{de}$ eventually turns to be $w_{de} < -1$ due to the expansion
of the universe. Thus the CHDE is responsible of the present cosmic accelerated expansion.

It is more interesting to observe that with the analytic feature of the differential equation of motion for $\Omega_{de}$, we are able to take the approximate
solution Eq.(\ref{frhom}) at certain point $a_{\rm i}\ll 1$ as the initial condition to solve the differential equation of motion for $\Omega_{de}$. Note that once
$d$ is given, the present fractional energy density $\Omega_{de}(a=1)$ can be naturally obtained by solving Eq.(\ref{ode}), so the degrees of freedom of the CHDE
model is the same as the one of the $\Lambda$CDM model.

Using the definition $a=1/(1+z)$ with $z$ the redshift, we can rewrite Eq. (\ref{ode}) as
 \begin{equation}
 \frac{d\Omega_{de}}{dz}=-\frac{\Omega_{de}(1-\Omega_{de})}{1+z}\left(11+\frac{r_0(1+z)}{1+r_0(1+z)}-\frac2d\sqrt{\Omega_{de}}(1+z)\right) ~. \label{odez}
 \end{equation}
Due to the analytical property mentioned above, we can take the approximate solution at $z_{i} \gg 1$ (or $a_i \ll 1$),
 \begin{equation}
 \Omega_{de}(z_{i})=\frac{d^2}{4}\left( 9+\frac{r_0(1+z_{i})}{1+r_0(1+z_{i})} \right)^2 \frac1{(1+z_{i})^2} ~,  \label{ini}
 \end{equation}
as the initial condition to solve the differential equation of motion for $\Omega_{de}$. The final solution depends weakly on the choice of $z_{i}$ in a wide range
as $\Omega_{de}$ is tiny and scales as $1/(1+z)^2$ at $z\gg 1$. This weak dependence is also checked directly by numerical method. In our numerical calculation, we
simply set $z_{i}=2000$.

\section{Best-fit Analysis on CHDE Model}

We will investigate the cosmological constraints on the CHDE model by using the Union2 compilation of 557 supernova Ia (SNIa) data \cite{Amanullah:2010vv}, the
baryon acoustic oscillation (BAO) results from the Sloan Digital Sky Survey data release 7 (SSDS DR7) \cite{Percival:2009xn} and  the cosmic microwave background
radiation (CMB) data from the 7-yr Wilkinson Microwave Anisotropy Probe (WMAP7) \cite{Komatsu:2010fb}. The analysis method for the observational data are given in
Appendix A.  In the following, we give the expression of the important quantity used in best-fit analysis, $E(z) \equiv H(z)/H_{0}$.

From the Friedmann equation Eq.(\ref{fri}), we have
\begin{equation}
E(z)\equiv {H(z)\over H_0}=\left(\Omega_{m0}(1+z)^3+\Omega_{r0}(1+z)^4\over 1-\Omega_{de}(z)\right)^{1/2}  ~, \label{Ez}
\end{equation}
At $z<2000$, the energy density of radiation is the sum of those of photons and relativistic neutrinos. Here $\Omega_{r0}=\Omega_{\gamma0}\left(1+0.2271N_{\rm
eff}\right)$, where $\Omega_{\gamma0}$ is the present fractional photon energy density and $N_{\rm eff}=3.04$ is the effective number of neutrino species
\cite{Komatsu:2010fb}. In this paper, we adopt the best-fit value, $\Omega_{\gamma0}=2.469\times10^{-5}h^{-2}$ (for $T_{CMB}=2.725$ K) with $h \equiv
H_{0}/100/[{\rm km} \, {\rm sec}^{-1} \, {\rm Mpc}^{-1}]$ given by WMAP7 \cite{Komatsu:2010fb}. For the given values of parameters $d$ and $r_0$, we can numerically
solve $\Omega_{de}(z) $ from the differential equation Eq.(\ref{odez}) by using the initial condition Eq.(\ref{ini}). Substituting the result $\Omega_{de}(z=0)$
into Eqs.(\ref{or1}) and (\ref{or2}), we obtain $\Omega_{r0}$ and $\Omega_{m0}$. With all the results, we then yield the function $E(z)$ from Eq.(\ref{Ez}).

\begin{table}
\caption{The best-fit results constrained by the observational data at 1 $\sigma$ (68.3\%) and 2 $\sigma$ (95.4\%) confidence levels}
\begin{center}
\label{fit}
\begin{tabular}{|c|ccc|}
\hline
                             & $\chi^2$  & ~  $ d $ ~                                         &  ~ $r_0 \times 10^4$                                 \\
\hline   $\Lambda$CDM        &  544.616  & ~  $ --                                        $   & ~    $3.057^{+0.100}_{-0.096}~ ^{+0.165}_{-0.157} $   \\
\hline     CHDE              &  546.273  & ~  $0.232^{+0.006}_{-0.006}~ ^{+0.009}_{-0.009}$   & ~    $3.052^{+0.074}_{-0.073}~ ^{+0.122}_{-0.120}$    \\
 \hline
\end{tabular}
\begin{tabular}{|c|cccc|}
\hline
                                   & ~                $\Omega_{m0}$                             & ~              $\Omega_{de0}$                             & ~                    $h$                            & ~ $w_{de0}$ (EOS)                                    \\
\hline  $\Lambda$CDM               & ~    $0.277^{+0.021}_{-0.019}~ ^{+0.035}_{-0.031}$         & ~    $0.723^{+0.019}_{-0.021}~ ^{+0.031}_{-0.035}$        & ~  $0.702^{+0.016}_{-0.015}~ ^{+0.026}_{-0.025}$    & ~   $             -1   $                             \\
\hline      CHDE                   & ~    $0.286^{+0.019}_{-0.018}~ ^{+0.032}_{-0.028}$         & ~    $0.714^{+0.018}_{-0.019}~ ^{+0.028}_{-0.032}$        & ~  $0.692^{+0.016}_{-0.015}~ ^{+0.026}_{-0.025}$    & ~   $  -1.240^{+0.027}_{-0.027}~ ^{+0.045}_{-0.044} $\\
\hline
\end{tabular}
\end{center}
\end{table}

\begin{figure}
\centerline{\includegraphics[width=15cm]{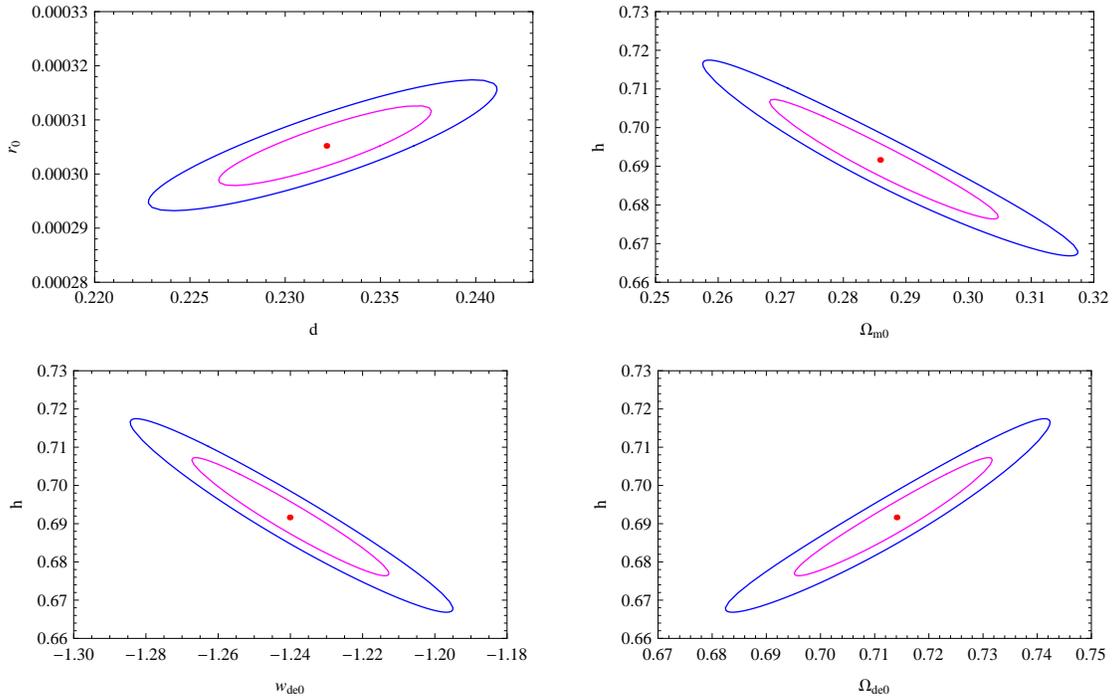}} \caption{\label{figfit}  Probability contours at 68.3\% and 95.4\% confidence levels for the CHDE model.}
\end{figure}

The best-fit results are summarized in TABLE \ref{fit}. For comparison, we also give the fitting results for the $\Lambda$CDM model with the same observational
data. Because of the tiny ratio of the fraction of radiation to that of matter $r_0\sim O(10^{-4})$, we have $\Omega_{m0}+\Omega_{de0}\simeq1$ in TABLE \ref{fit}.
It is seen that the CHDE model favors slightly larger fraction of matter. Obviously, the present EoS of CHDE is below $-1$ significantly. In Fig. \ref{figfit}, we
plot some probability contours at 68.3\% and 95.4\% confidence levels for the relevant cosmological quantities in the CHDE model.

\section{Systematic study on CHDE model}

Taking the best-fit values of model parameters, we are able to investigate the cosmic evolutions of some interesting quantities by solving corresponding equations
numerically. Here we are going to briefly study the cosmic evolutions of the fractional energy density $\Omega_{de}$, the equation of state $w_{de}$, the
deceleration parameter and statefinder \cite{Chiba:1998,Sahni:2003} in the CHDE model.

\begin{figure}
\centerline{\includegraphics[width=14cm, height=8cm]{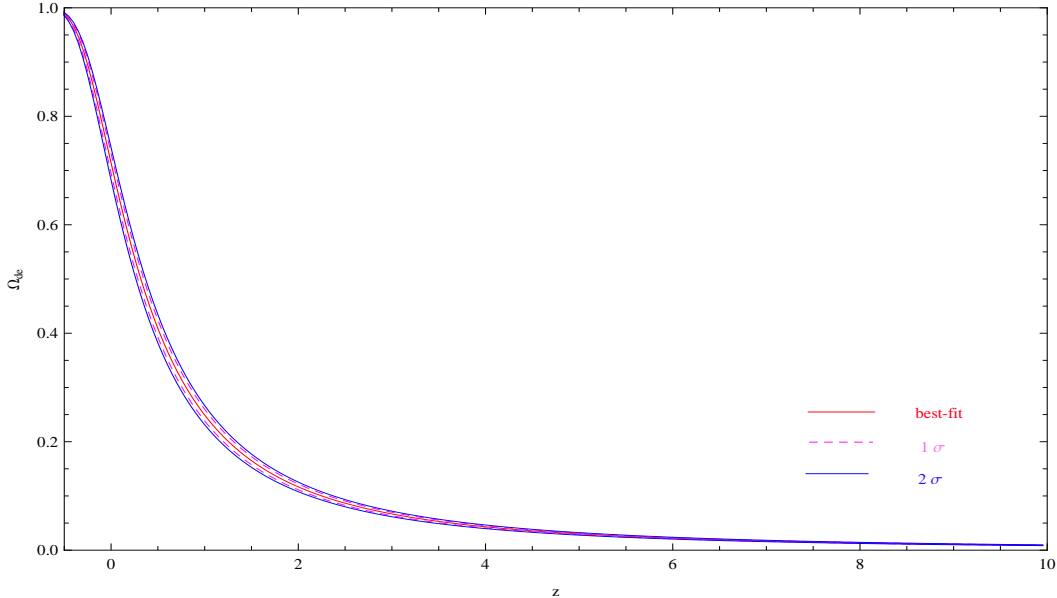}} \caption{\label{figol}  Cosmic evolution of the fractional energy density of CHDE}
\end{figure}
\begin{figure}
\centerline{\includegraphics[width=14cm, height=8cm]{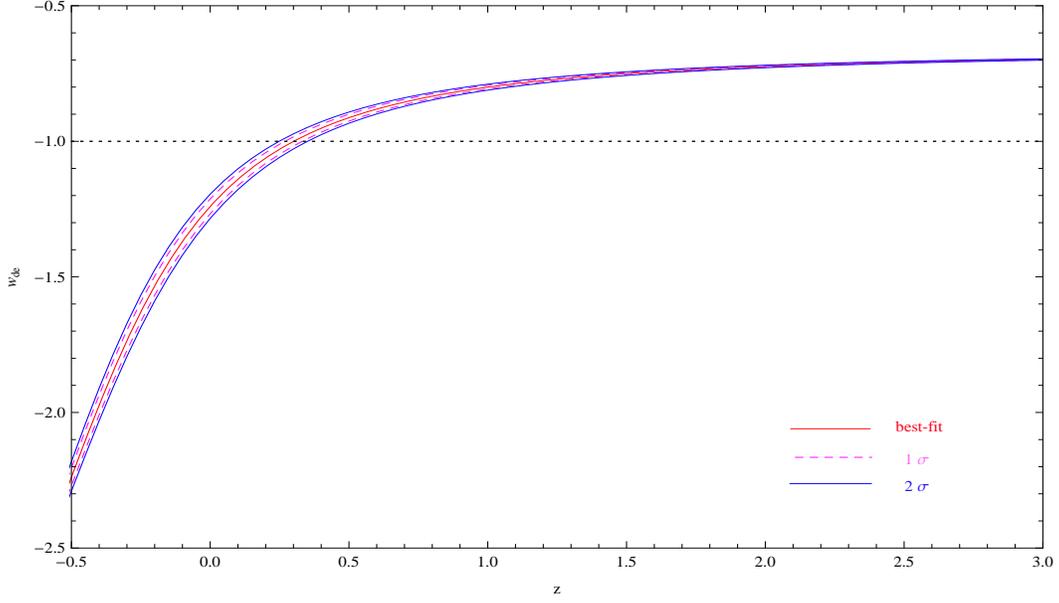}} \caption{\label{figwde}  Cosmic evolution of the EoS of CHDE}
\end{figure}
\begin{figure}
\centerline{\includegraphics[width=14cm, height=8cm]{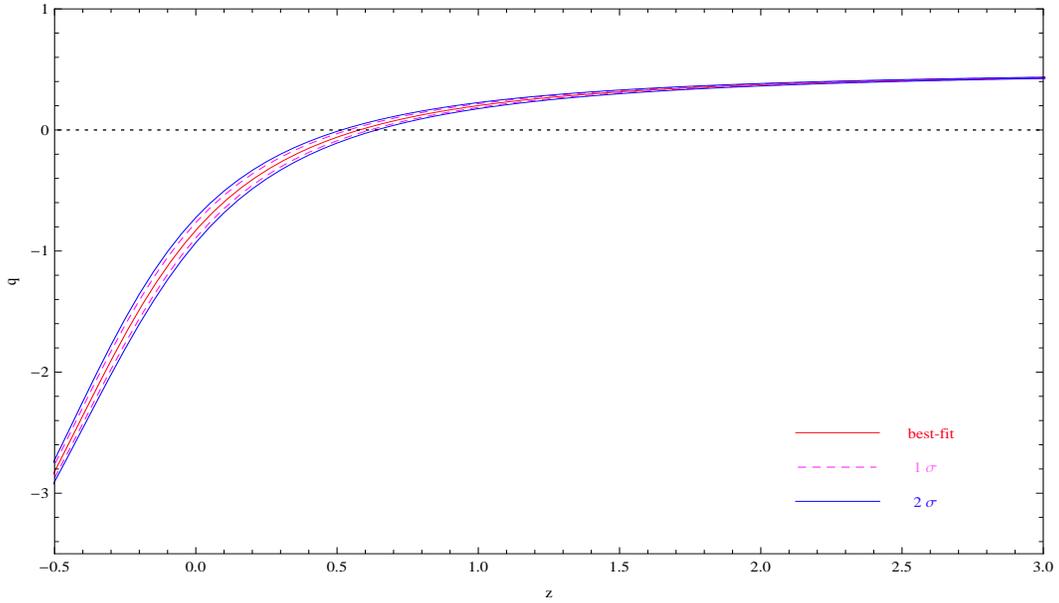}} \caption{\label{figq} Eolution of the decelerating parameter}
\end{figure}

In Fig.\ref{figol}, we present the evolutionary trajectory of the fractional energy density of CHDE. It is seen that the fractional energy density of CHDE decreases
rapidly with increase of the redshift and becomes tiny in the early universe, thus the model is consistent with primordial nucleosynthesis (BBN)\cite{Olive:1999ij}.
Actually, this can be enlightened by the fact from Eq.(\ref{frhom}) that $\Omega_{de}\propto a^{2}$ with the proportionality coefficient in order of $O(1)$ by
referring to the best-fit results of the model parameters. Obviously, $\Omega_{de}\ll1$ at $a \ll 1$ in the early universe.

Substituting the results of $\Omega_{de}$ to Eq.(\ref{wde}), we get the equation of state of CHDE. The evolutionary trajectory of the EoS of CHDE is shown in
Fig.\ref{figwde}. It is observed that the EoS of CHDE cross over $w_{de}=-1$ from $w_{de}>-1$ to $w_{de}<-1$ during the universe expansion. The potential
singularity for $w_{de}<-1$ in the future might be ceased by high order effect of gravity.

From the conservation of total energy $\dot{\rho}_{tot}+3H(1+w_{tot})\rho_{tot}=0$, we have $w_{tot}=-1-\frac23\frac{\dot{H}}{H^2}$ by using $\rho_{tot}=3M_p^2H^2$.
On the other hand, $w_{tot}=\frac13 \Omega_r+w_{de}\Omega_{de}$ as matter is pressureless. Combining with Eq.(\ref{or2}), it is not difficult to get the
decelerating parameter
 \begin{equation}
    q \equiv-\frac{\ddot{a}}{aH^2}\equiv -1-\frac{\dot{H}}{H^2}=\frac12+\frac12\frac{r_0(1+z)}{1+r_0(1+z)}(1-\Omega_{de})+\frac32w_{de}\Omega_{de} ~.\label{q}
 \end{equation}
Using the expression of $w_{de}$ in Eq.(\ref{wde}) and substituting the results of $\Omega_{de}$ into above equation, we then yield the decelerating parameter. The
evolutionary trajectory of the decelerating parameter is shown in Fig.\ref{figq}. It is clear that the universe does transit from the decelerated expansion $q>0$ to
the accelerated expansion $q<0$ at the very recent epoch.

\begin{figure}
\centerline{\includegraphics[width=10cm, height=12cm]{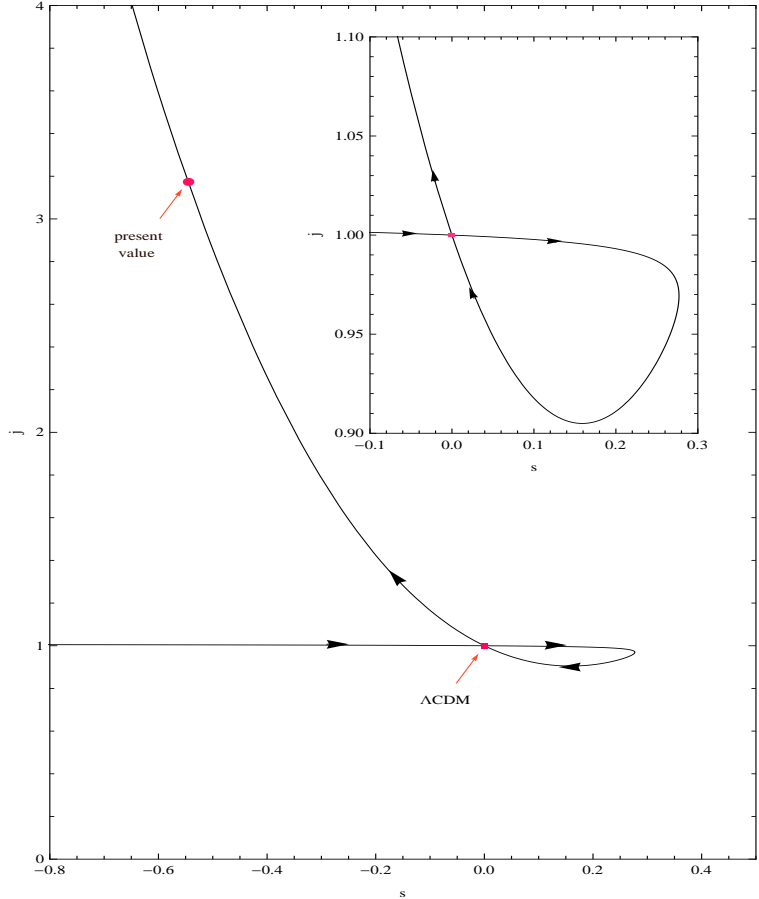}} \caption{\label{figjs}  The statefinder parameter $j-s$ contour evolves in redshift inteval $z \in
[-0.2,15]$ (the arrow indicates the evolution from high redshift to low redshift); where model parameters take the best-fit values, i.e. $d=0.232$ and
$r_0=3.052\times 10^{-4}$}
\end{figure}

The statefinder \cite{Chiba:1998,Sahni:2003} are geometric parameters probing the expansion dynamics of the universe through high derivatives of the scale factor
$\ddot a$ and $\dddot a$.  It is a natural next step beyond the Hubble parameter $H$ depending on $\dot a$ and the deceleration parameter $q$ depending on $\ddot
a$. The statefinder pair $\{j, s\}$ are defined as
 \begin{equation}
    j\equiv\frac{\dddot a}{aH^3} ~, \label{j1}
 \end{equation}
 \begin{equation}
   s\equiv\frac{j-1}{3(q-\frac{1}{2})} ~, \label{s1}
 \end{equation}
where we use $j$ instead of $r$ used in \cite{Sahni:2003} to denote the first parameter because we have used $r$ to denote the ratio of the fraction energy density
of radiation to that of matter defined in Eq.(\ref{or1}).

The $\Lambda$CDM model corresponds to a fixed point $\{ j, s \} = \{1,0\} $. Departure of a given dark energy model from this fixed point provides a good way of
establishing the "distance" of this model from the $\Lambda$CDM model \cite{Sahni:2003}.

Using the FRW equation Eq.(\ref{fri}), conservation equations Eqs.(\ref{ceq}) and the decelerating parameter Eq.(\ref{q}), we can rewrite the satefinder parameters
as
 \begin{eqnarray}
    j&=&1+\frac32\left( (1+z)\frac{dw_{de}}{dz} + 3w_{de}(1+w_{de})   \right) \Omega_{de}+2\Omega_r ~, \\
    s&=& \frac{3\left( (1+z)\frac{dw_{de}}{dz} + 3w_{de}(1+w_{de})   \right) \Omega_{de}+4\Omega_r}{9w_{de}\Omega_{de}+3\Omega_r}  ~.  \label{s2}
 \end{eqnarray}
Referring to Eq.(\ref{wde}) and Eq.(\ref{or2}), $w_{de}$ and $\Omega_r$ can be expressed in terms of $\Omega_{de}$. Therefore, solving the differential equation of
$\Omega_{de}(z)$ and substituting the results into above equations, the statefinder parameter pair $\{j(z), s(z)\}$ can be obtained.

As the present fraction of radiation is tiny and the EoS of dark energy is negative at present epoch, the denominator of $s$ in Eq.(\ref{s2}) is negative today.
However, the fraction of radiation energy increases rapidly with increase of the redshift while the fraction of dark energy is tiny in the early universe. Thus, the
denominator is positive in the early universe. Therefore, the statefinder parameter $s$ might be divergent at some early time and would become meaningless. In
Fig.[\ref{figjs}], we only show the evolutionary trajectory of the statefinder parameter $j-s$ in redshift interval $z \in [-0.2,15]$. The present statefinder of
the CHDE model is significantly away from the fixed point corresponding to the $\Lambda$CDM model. Thus, statefinder may be a sensitive diagnostic to differentiate
these two models.

\section{Concluding Remarks}

We have made the best-fit analysis on the holographic dark energy model characterized by the conformal-age-like length $L=\frac1{a^4(t)}\int_0^t dt'~a^3(t')$ (CHDE)
\cite{Huang:2012nz}.  With the joint analysis by using the Union2 compilation of 557 supernova Ia (SNIa) data \cite{Amanullah:2010vv}, the baryon acoustic
oscillation results from SSDS DR7 \cite{Percival:2009xn} and the cosmic microwave background radiation (CMB) data from the WMAP7 \cite{Komatsu:2010fb}, we have
obtained the minimal $\chi^2$ for CHDE model with $\chi^2_{min}=546.273$, which is slightly bigger than the one for the $\Lambda$CDM model with
$\chi^2_{min}=544.616$. The constraints on the model parameter $d$ at 1 $\sigma$ and 2 $\sigma$ confidence levels are found to be $d=0.232\pm 0.006\pm 0.009$. Corresponding constraints on the present fractional energy density of matter and the present equation of state of CHDE are found to be
$\Omega_{m0}=0.286^{+0.019}_{-0.018}~ ^{+0.032}_{-0.028}$ and $w_{de0}=-1.240^{+0.027}_{-0.027}~ ^{+0.045}_{-0.044}$ respectively. For comparison, we have also
fitted the $\Lambda$CDM model to the same observational data and found that $\Omega_{m0}=0.277^{+0.021}_{-0.019}~ ^{+0.035}_{-0.031}$. It has been seen that the
CHDE model leads to a slightly larger fraction of matter comparing to the $\Lambda$CDM model and the present EoS of CHDE is much less than $-1$.

We have also provided a systematic analysis on the CHDE model for its cosmic evolutions of the fractional energy density of dark energy, the EoS of dark energy, the
deceleration parameter and statefinder. From the evolutionary trajectory of $\Omega_{de}$, it has been found that the fraction of dark energy decreases rapidly with
the increase of the redshift and becomes tiny in the early universe. Thus, the model is consistent with the primordial nucleosynthesis (BBN) \cite{Olive:1999ij}.
The EoS of CHDE has been shown to cross from $w_{de}>-1$ to $w_{de}<-1$ during the universe expansion. The evolution of the deceleration parameter has indicated
that the universe transits from decelerated expansion $q>0$ to accelerated expansion $q<0$ recently. It has been noticed that the statefinder may provide a
sensitive diagnostic to differentiate the CHDE model with the $\Lambda$CDM model.

\section*{Acknowledgements}
We would like to thank Miao Li for useful discussions. The author (Z.P.H) would like to thank M. Q. Huang and M. Zhong for their helpful support. This work is
supported in part by the National Basic Research Program of China (973 Program) under Grants No. 2010CB833000; the National Nature Science Foundation of China
(NSFC) under Grants No. 10975170, 10975184, 10947016.

\appendix

\section{ Analysis method for the observational data}

In this appendix, we explain the methods for the elaboration of observational data from the Union2 compilation of 557 supernova Ia (SNIa) data
\cite{Amanullah:2010vv}, the baryon acoustic oscillation (BAO) results from the Sloan Digital Sky Survey data release 7 (SSDS DR7) \cite{Percival:2009xn} and the
cosmic microwave background radiation (CMB) data from the 7-yr Wilkinson Microwave Anisotropy Probe (WMAP7) \cite{Komatsu:2010fb}. For the three independent
observations, the likelihood function of a joint analysis is given by
 \begin{eqnarray}
  \cal{L } &=& \cal{L }_{\rm SN} \times \cal{L }_{\rm BAO} \times \cal{L }_{\rm CMB} \nonumber \\
    &=& \exp[-(\tilde{\chi}^2_{\rm SN}+\chi_{\rm BAO}^2+\chi_{\rm CMB}^2)/2]~.
 \end{eqnarray}
The model parameters yielding a maximal $\cal{L }$, thus a minimal $\chi^{2}=\tilde{\chi}_{\rm SN}^{2}+\chi_{\rm BAO}^{2}+\chi_{ \rm CMB}^{2}$, will be favored by
the observations. In the following, we present the calculation for the various  $\chi^{2}_i$ of each observational data set, and mainly adopt the analysis method
described in \cite{Geng:2011ka,Bamba:2010wb,Yang:2010xq}.

\subsubsection{ Type Ia Supernovae (SN Ia)}

The SN Ia observations give the information on the luminosity distance $D_{L}$. The distance modulus is theoretically defined as the function of the redshift $z$
\begin{equation}
\mu_{\rm th}(z_i)\equiv 5 \log_{10} {D_L(z_i)} +\mu_0   ~,
\end{equation}
with $\mu_{0}\equiv42.38-5\log_{10}h$ and $h \equiv H_{0}/100/[{\rm km} \, {\rm sec}^{-1} \, {\rm Mpc}^{-1}]$. The Hubble-free luminosity distance has the following
form for the flat universe
\begin{equation}
 D_{L}(z)=\left(1+z\right)\int_{0}^{z}\frac{dz'}{E(z')} ~,
\end{equation}
with $E(z) \equiv H(z)/H_{0}$.

The $\chi^2$ for the SNIa data is given by
\begin{equation}
\chi^2_{\rm SN}=\sum\limits_{i}{[\mu_{\rm obs}(z_i)-\mu_{\rm th}(z_i)]^2\over \sigma_i^2} ~, \label{ochisn}
\end{equation}
where $\mu_{\rm obs}(z_i)$ and $\sigma_i$ are the observed quantity and the corresponding 1$\sigma$ error of distance modulus for each supernova, respectively.
Adopting the approach in \cite{Perivolaropoulos:2004yr}, the $\chi^2_{\rm SN}$  with respect to $\mu_0$ can be expanded as
\begin{equation}
\chi^2_{\rm SN}=A-2\mu_0 B+\mu_0^2 C ~, \label{ochisn2}
\end{equation}
with
\begin{equation}
A=\sum\limits_{i}{[\mu_{obs}(z_i)-\mu_{th}(z_i;\mu_{0}=0)]^2\over \sigma_i^2}  ~,
\end{equation}
\begin{equation}
B=\sum\limits_{i}{\mu_{obs}(z_i)-\mu_{th}(z_i;\mu_{0}=0)\over \sigma_i^2} ~,
\end{equation}
\begin{equation}
C=\sum\limits_{i}{1\over \sigma_i^2} ~.
\end{equation}
It is easy to check that the minimum of $\chi_{\mathrm{SN}}^{2}$ with respect to $\mu_{0}$ is given by
\begin{equation}
\tilde{\chi}^2_{\rm SN}=A-\frac{B^2}C ~. \label{tchi2sn}
\end{equation}
which is applied in our best-fit analysis to the $\chi^{2}$ minimization by using the Supernova Cosmology Project (SCP) Union2 compilation, which contains 557
supernovae~\cite{Amanullah:2010vv} with the range of the redshift $z=0.015-1.4$.

\subsubsection{Baryon Acoustic Oscillations (BAO)}

The distance ratio $d_z\equiv r_s(z_d)/D_V(z)$ is measured by BAO observations, with $r_{s}$ the comoving sound horizon, $D_{V}$ the volume-averaged distance and
$z_d$ the redshift of the baryon drag epoch \cite{Percival:2009xn}.

The comoving sound horizon $r_{s}(z)$ is given by
\begin{equation}
 r_s(z) = {1\over \sqrt{3}} \int_{0}^{1/(1+z)} {da \over a^2H(a)\sqrt{1+(3\Omega_{b0}/4\Omega_{\gamma0})a} }  ~, \label{rs}
\end{equation}
with $\Omega_{b0}$ and $\Omega_{\gamma0}$ corresponding to the present baryon and photon density parameters. We take the best-fit values: $\Omega_{b0}=0.02253
h^{-2}$ and $\Omega_{\gamma0}=2.469\times10^{-5}h^{-2}$ (for $T_{CMB}=2.725$ K) obtained by the 7-yr WMAP observations \cite{Komatsu:2010fb}.

The volume-averaged distance $D_{V}(z)$ is defined  as \cite{Eisenstein:2005su}
\begin{equation}
 D_{V}(z)\equiv\left[\left(1+z\right)^{2}D_{A}^{2}(z)\frac{z}{H(z)}\right]^{1/3} ~,
\end{equation}
with $D_{A}(z)$ the proper angular diameter distance, which is defined for the flat universe by
\begin{equation}
 D_{A}(z) \equiv \frac{1}{1+z}\int_{0}^{z}\frac{dz'}{H(z')}~.
\end{equation}
and $z_d$ is given by \cite{Eisenstein:1997ik}
\begin{equation}
 z_d = {1291 (\Omega_{m0}h^2)^{0.251}\over 1+0.659(\Omega_{m0}h^2)^{0.828}}\left[1+b_1(\Omega_{b0}h^2)^{b_2}\right] ~,
\end{equation}
with
\begin{equation}
 b_1 = 0.313 (\Omega_{m0}h^2)^{-0.419}\left[1+0.607(\Omega_{m0}h^2)^{0.674}\right], \;\;\; b_2 = 0.238(\Omega_{m0}h^2)^{0.223} ~.
\end{equation}

From the Two-Degree Field Galaxy Redshift Survey (2dFGRS) and the Sloan Digital Sky Survey Data Release 7 (SDSS DR7) \cite{Percival:2009xn}, we have the values for
the distance ratio $ d^{\rm obs}_{0.2} = 0.1905$ and $d^{\rm obs}_{0.35} = 0.1097 $ corresponding to the two redshifts $z=0.2$ and $z=0.35$. The $\chi^2$ of the BAO
data is given by:
\begin{equation}
 \chi^2_{\rm BAO} = Y^{\rm T}C_{BAO}^{-1}Y ~,
\end{equation}
with $ Y=\left(\begin{array}{cc} d^{\rm th}_{0.2}-d^{\rm obs}_{0.2}~, & ~d^{\rm th}_{0.35}-d^{\rm obs}_{0.35} \end{array}\right)^{\rm T}$, and the inverse
covariance matrix
\begin{equation}
C_{\rm BAO}^{-1}=\left(
\begin{array}{cc}
30124 & -17227\\
-17227 & 86977
\end{array}\right) ~ .
\end{equation}

\subsubsection{Cosmic Microwave Background (CMB)}

For the CMB data, we shall use the acoustic scale $l_A$, the shift parameter $R$ and the redshift of the decoupling epoch of photons $z_{*}$. They are defined as
\cite{Bond:1997wr}
\begin{eqnarray}
    l_{A}(z_{*}) &\equiv& \left(1+z_{*}\right)\frac{\pi D_{A}(z_{*})}{r_{s}(z_{*})} ~,\\
     R(z_{*}) &\equiv& \sqrt{\Omega_{m0}}H_{0} \left(1+z_{*}\right)D_{A}(z_{*}) ~,
\end{eqnarray}
with $r_{s}$ given in Eq.(\ref{rs}). The redshift of the decoupling epoch $z_{*}$ is given by~\cite{Hu:1995en}
\begin{equation}
 z_{*}=1048[1+0.00124(\Omega_{b0} h^2)^{-0.738}][1+g_1(\Omega_{m0} h^2)^{g_2}] ~,
\end{equation}
with
\begin{equation}
g_1=\frac{0.0783(\Omega_{b0} h^2)^{-0.238}}{1+39.5(\Omega_{b0} h^2)^{0.763}},\quad g_2=\frac{0.560}{1+21.1(\Omega_{b0} h^2)^{1.81}} ~.
\end{equation}

From WMAP7 observations \cite{Komatsu:2010fb}, we have the corresponding CMB data $l^{\rm obs}_{A}(z_{*})=302.09$, $R^{\rm obs}(z_{*})=1.725$ and $z^{\rm
obs}_{*}=1091.3$. The resulting $\chi^{2}$ is
\begin{equation}
 \chi^2_{\rm CMB} = X^{\rm T}C_{CMB}^{-1}X  ~,
\end{equation}
with $ X=\left(\begin{array}{ccc} l^{\rm th}_{A}(z_{*})-l^{\rm obs}_{A}(z_{*})~, &~ R^{\rm th}(z_{*})-R^{\rm obs}(z_{*})~, &~ z^{\rm th}_{*}-z^{\rm obs}_{*}
\end{array}\right)^{\rm T}$, and the inverse covariance matrix
\begin{equation}
C_{\rm CMB}^{-1}=\left(\begin{array}{ccc}
2.305 & 29.698 & -1.333\\
29.698 & 6825.27 & -113.180\\
-1.333 & -113.180 & 3.414\end{array}\right)  ~.
\end{equation}
%


\end{document}